\documentclass[12pt,a4paper]{article}
\usepackage{amssymb}
\usepackage{amscd}
\usepackage{amsthm}

\newcommand{\comments}[1]{}
\newcommand{\deff}{\emph}
\newcommand{\R}{\mathbb{R}}

\newcommand{\OmegaEL}{\Omega_{\mathrm{EL}}}

\newcommand{\HEL}{H_{\mathrm{EL}}}
\newcommand{\HdR}{H_{\mathrm{dR}}}
\newcommand{\XS}{\mathcal{X}_{\mathrm{S}}}
\newcommand{\XH}{\mathcal{X}_{\mathrm{H}}}
\newcommand{\boldgrk}[1]{\mbox{\boldmath $#1$}}
\newcommand{\vect}{\boldgrk}
\newcommand{\dd}{\mathrm{d}}
\newcommand{\tr}{\mathrm{tr\,}}
\newcommand{\im}{\mathrm{im}}
\newcommand{\Lied}[1]{\mathcal{L}_{\vect{#1}}}
\newcommand{\at}[1]{|_{#1}}

\newcommand{\h}{\hat{h}}
\newcommand{\e}{\hat{e}}
\newcommand{\f}{\hat{f}}

\newcommand{\Span}{\mathrm{span}}
\newcommand{\omits}[1]{}

\newtheorem{thm}{Theorem}
\newtheorem{corol}{Corollary}[thm]
\newtheorem{lem}{Lemma}
\newtheorem{proposition}{Proposition}

\newcounter{defc}

\newcounter{remkc}

\begin{document}

 \title{\bf The Euler-Lagrange
Cohomology Groups
  \\ on Symplectic Manifolds}

\author{Han-Ying Guo\thanks{Email: hyguo@itp.ac.cn}\\
  CCAST-WL, P. O. Box 8730, Beijing 100080, China. \\[1mm]
  Institute of Theoretical Physics, Chinese Academy of Sciences, \\
  P. O. Box 2735, Beijing 100080, China.\\[1mm]
  Jianzhong Pan\thanks{Email: pjz@mail.amss.ac.cn}\\
Institute of Mathematics, Chinese Academy of Sciences\\
   Beijing 100080, China.\\[1mm]
      Ke Wu\thanks{Email: wuke@itp.ac.cn}\\
  Department of Mathematics, Capital Normal University, \\
  Beijing 100037, China.\\[1mm]
 \omits{ Siye Wu\thanks{Email: Siye.Wu@colorado.edu}\\
Department of Mathematics, Colorado University, Colorado,
USA.\\[1mm]}
  and\\[1mm]
   Bin Zhou\thanks{Email: zhoub@ihep.ac.cn}\\
  Institute of High Energy Physics, Chinese Academy of Sciences, \\
  P. O. Box 918-4, Beijing 100039, China.}


\maketitle

\begin{abstract}
The definition and properties of the Euler-Lagrange co\-homology
groups $\HEL^{2k-1}$, $1 \leqslant k \leqslant n$, on a symplectic
manifold $({\cal M}^{2n},\omega)$ are given and studied. For $k =
1$ and $k = n$, they are isomorphic to the corresponding de~Rham
cohomology groups $\HdR^1({\cal M}^{2n})$ and $\HdR^{2n-1}({\cal
M}^{2n})$, respectively. The other Euler-Lagrange cohomology
groups are different from either the de~Rham cohomology groups or
the harmonic cohomology groups on $({\cal M}^{2n},\omega)$, in
general. The general volume-preserving equations on $({\cal
M}^{2n},\omega)$ are also presented from cohomological point of
view. In the special cases, these equations become the ordinary
canonical equations in the Hamilton mechanics. Therefore, the
Hamilton mechanics has been generalized via the cohomology.
\end{abstract}

\newpage \tableofcontents

\newpage
\section{Introduction} \label{sect:Intro}

It is well known that the theory on symplectic manifolds plays an
important role in both classical mechanics (see, for example,
\cite{Arnold}, \cite{AM}) and field theory. On the other hand,
both Lagrange and Hamilton mechanics had been also well
established.

Very recently, however, the Euler-Lagrange cohomology has first
been introduced and discussed in \cite{ELcoh2,ELcoh4} for
classical mechanics and  field theory in order to explore the
relevant topics in\omits{. They were discussed in both Lagrange
and Hamilton formalism as well as their} the (independent
variable(s)) discrete mechanics and field theory including
symplectic and multisymplectic algorithms. \omits{Here we want to
present a through discussion on those for the classical mechanics,
in both the Lagrange and Hamilton formalism.}

Based upon these works, we have further found that there is in
fact a sequence of the particular cohomology groups called the
Euler-Lagrange cohomology groups on the symplectic manifolds. What
has been found in \cite{ELcoh2,ELcoh4} in the case of the
classical mechanics is the (first) Euler-Lagrange cohomology group
of the Euler-Lagrange $1$-forms. We have also found that these
cohomology groups may play some important role in the classical
mechanics as well as other dynamical systems such as the
volume-preserving systems and so on (some aspects on these issues
have been given in \cite{zgw,  gpwz, zgp}).

In this paper, we introduce the general definition of these
Euler-Lagrange cohomology groups $\HEL^{2k-1}({\cal
M}^{2n},\omega)$, $1 \leqslant k \leqslant n$, on a
$2n$-dimensional symplectic manifold $({\cal M}^{2n},\omega)$ with
the symplectic structure $\omega$ and study their properties in
some details. We show that for $k = 1$ and $k = n$, they are
isomorphic to the corresponding de~Rham cohomology groups
$\HdR^1({\cal M}^{2n})$ and $\HdR^{2n-1}({\cal M}^{2n})$,
respectively. Consequently, due to the Poincar\'e duality, the
first Euler-Lagrange cohomology group $\HEL^{1}({\cal
M}^{2n},\omega)$ and the
 highest one $\HEL^{2n-1}({\cal M}^{2n},\omega)$ are dual to each
other.  We also show that the other Euler-Lagrange cohomology
groups $\HEL^{2k-1}({\cal M}^{2n},\omega), ~1<k<n,$ are different
from either the de~Rham cohomology groups or the harmonic
cohomology groups on $({\cal M}^{2n},\omega)$, in general.

From the cohomological point of view, the ordinary Hamiltonian
canonical equations correspond to 1-forms that represent trivial
element in  the first Euler-Lagrange group on $({\cal M}^{2n},\omega)$
as the phase space. Analog to this fact, it is natural but significant
to find the general volume-preserving equations on $({\cal M}^{2n},\omega)$
from such forms that represent trivial element in the highest Euler-Lagrange
cohomology group $\HEL^{2n-1}({\cal M}^{2n},\omega)$. In this paper, we
introduce this general kind of volume-preserving equations from this
point of view. In the special cases, these equations become the ordinary
canonical equations in the Hamilton mechanics. Therefore, the Hamilton
mechanics has been generalized to the volume-preserving systems on
symplectic manifolds via the cohomology.

This paper is arranged as follows. In section 2, we first briefly
recall the definition of the first Euler-Lagrange cohomology group
on a symplectic manifold $({\cal M}^{2n},\omega)$ and prove that
it is isomorphic to the first de Rahm cohomology group on the
manifold. Then we introduce the general definition of the $2k-1$st
Euler-Lagrange cohomology groups for $1\leqslant k \leqslant n$ on $({\cal
M}^{2n},\omega)$ in section 3. We first indicate that the highest one
is equivalent to the $2n-1$st de Rahm cohomology group. We also indicate
that in general they are not isomorphic to each other and that they are not
isomorphic to either the de Rahm cohomology or the harmonic
cohomology on $({\cal M}^{2n},\omega)$. In section 4, the relative
Euler-Lagrange cohomology is introduced in analog with the
relative de Rahm cohomology. The general volume-preserving
equation is introduced in section 5. Its relations with ordinary
canonical equations in the Hamilton mechanics as well as other
volume-preserving systems are discussed. It is clear that the
general volume-preserving equations  are the generalization of the
ordinary canonical equations in Hamilton Mechanics. Finally, we
end with some discussion and remarks in section 6.

\section{The First Euler-Lagrange Cohomology  Group on Symplectic  Manifolds}
\label{sectELsymp}

In order to set up notations, we briefly introduce the first
Euler-Lagrange cohomology group for  what are called the
Euler-Lagrange $1$-forms on $({\cal M},\omega)$ in this
subsection. We also prove that it is isomorphic to the first
de~Rham cohomology group on it.

For a Hamiltonian system with Hamiltonian $H=H(q,p)$ on a
$2n$-dimensional symplectic manifold $({\cal M}, \omega)$, the
trajectory, $q=q(t)$ and $p=p(t)$, is determined by the Hamilton
principle. Namely, it is a stationary point of the action
functional
\begin{displaymath}
  S[q(t),p(t)] := \int_{t_0}^{t_1}\{\,p_i(t)\dot{q}^{\,i}(t)
  - H(q(t),p(t))\,\}\,
  \dd t
\end{displaymath}
with $\delta q^i(t_0)=\delta q^i(t_1) = 0$ provided that $q=q(t)$ and
$p=p(t)$ satisfy the canonical equations
\begin{equation}\label{Heqn}
  \dot{q}^{\,i} = \frac{\partial H}{\partial p_i}, \qquad
  \dot{p}_i = - \frac{\partial H}{\partial q^i}.
\end{equation} \omits{It is similar that we
should consider arbitrary curves on ${\cal M}$. Thus, similarly to
the case in the Lagrangian mechanics, }For arbitrary curves on
${\cal M}$, the \deff{Euler-Lagrange 1-form}
\begin{equation}
  E := \Big(\dot{q}^{\,i} - \frac{\partial H}{\partial p_i}\Big)\,\dd p_i
    - \Big(\dot{p}_i + \frac{\partial H}{\partial q^i}\Big)\,\dd q^i
  = \dot{q}^{\,i}\,\dd p_i - \dot{p}_i\,\dd q^i - \dd
  H
\label{ELham} \end{equation}  can be defined along a curve and
$E=0$ gives rise to the canonical equations (\ref{Heqn}).

In fact, it can be defined on the whole manifold $\mathcal{M}$ by
introducing a congruence of maximal integral curves of a smooth
vector field on ${\cal M}$
\begin{equation}
  \vect{X} = \dot{q}^{\,i}(q,p)\,\frac{\partial}{\partial q^i}
  + \dot{p}_i(q,p)\,\frac{\partial}{\partial p_i},
\label{Xdef}
\end{equation}
where $\dot{q}^{\,i} = \dot{q}^{\,i}(q,p)$ and $\dot{p}^{\,i} =
\dot{p}^{\,i}(q,p)$ are functions on ${\cal M}$.\omits{: the
components of the vector field relative to the given congruence of
curves.}

Given such a vector field,  the Euler-Lagrange 1-form can be
constructed as follows: The exterior differential of the function
$p_i\dot{q}^{\,i} - H$,
\begin{eqnarray*}
  \dd(p_i\dot{q}^{\,i} - H)
  =\dot{q}^{\,i}\,\dd p_i - \dot{p}_i\,\dd q^i - \dd H
   + \frac{\dd}{\dd t}(p_i\,\dd q^i),
\end{eqnarray*} leads to \begin{equation}
  \dd(p_i\dot{q}^{\,i} - H) = E + \frac{\dd\theta}{\dd t}
\label{dH} \end{equation} where \begin{equation}
  \theta := p_i\,\dd q^i.
\end{equation} In the above calculation, it is
used that $\dd$ commutes  with $\frac{\dd}{\dd t}$. This is
because $\frac{\dd}{\dd t}$ along any one of these integral curves
is nothing but the restriction of the Lie derivative $\Lied{X}$.
While $\Lied{X}$ commutes with $\dd$, so does $\frac{\dd}{\dd t}$.

It is clear that the Euler-Lagrange 1-form (\ref{ELham}) depends
on the Hamiltonian function $H$ and a smooth vector field
$\vect{X}$ as in eq.~(\ref{Xdef}). \omits{ Hence $\dot{q}^{\,i} =
\Lied{X}q^i$ and
 $\dd\dot{q}^{\,i} = \dd\Lied{X}q^i=\Lied{X}\dd q^i = \frac{\dd}{\dd t}\dd q^i$,
because $\Lied{X} = \dd i_{\vect{X}} + i_{\vect{X}}\dd$ is
commutative with $\dd$.

When $\mathcal{M}=T^*{M}$ is the cotangent bundle of the
configuration manifold $M$, both $p_i\dot{q}^{\,i}$ and $p_i\,\dd
q^i$ can be globally defined on $\mathcal{M}$, hence
eq.~(\ref{dH}) is a globally defined decomposition. While for a
generic symplectic manifold $(\mathcal{M},\omega)$, the Darboux
coordinates $q^i$ and $p_i$ are merely locally defined functions.
In this case both $p_i\dot{q}^{\,i}$ and $p_i\,\dd q^i$ are
locally defined: They can not be ``glued" patch by patch smoothly
into smooth objects on $\mathcal{M}$. However} Furthermore, it can
be shown that the Euler-Lagrange 1-form $E$ is globally
defined.\omits{ done so, and eq.~(\ref{dH}) is always valid on
every Darboux chart. }

Due to the nilpotency of $\dd$, the second operation of $\dd$ on
(\ref{dH}) leads to the globally valid formula \begin{equation}
  \dd E = - \frac{\dd\omega}{\dd t}
\end{equation} \omits{when we take the exterior
differentials of both sides of eq.~(\ref{dH}),} where
\begin{equation}\label{omega}
  \omega := \dd\theta = \dd p_i \wedge \dd q^i
\end{equation} is the symplectic form on
$\mathcal{M}$. Thus, we have
\begin{thm}
  The symplectic form $\omega$ is conserved if and only if the Euler-Lagrange
1-form $E$ is closed.
\end{thm}

Let $\OmegaEL(\mathcal{M})$ denote the linear space (an Abelian
group)  generated by those Euler-Lagrange 1-forms. That is, a
1-form, say, $\alpha$  is in $\OmegaEL(\mathcal{M})$ if and only
if there exist finitely many Euler-Lagrange 1-forms $E_1$, \ldots,
$E_k$ so that
  $\alpha = E_1 + \ldots + E_k$.
It is easy to verify that $\OmegaEL(\mathcal{M})$ is a real linear
space.

Denote $Z_{\mathrm{ EL}}(\mathcal{M}):=\{\textrm{closed\ 1-forms\
in\ }\OmegaEL(\mathcal{M})\}$ and $B_{\mathrm{
EL}}(\mathcal{M}):=\{\textrm{exact\ 1-forms\ in\
}\OmegaEL(\mathcal{M})\}$. The quotient linear space (also a
quotient Abelian group)
\begin{displaymath}
  H_{\mathrm{EL}}(\mathcal{M},\omega)
  := Z_{\mathrm{
EL}}(\mathcal{M})
  /B_{\mathrm
{EL}}(\mathcal{M}) \end{displaymath} is called
\deff{the first Euler-Lagrange cohomology group}.

\omits{Recall that $\omega$, as a symplectic form on
$\mathcal{M}$, is a closed 2-form that is non-degenerate at every
point in $\mathcal{M}$. This implies that,}On the other hand, for
an arbitrary vector field $\vect{X}$, the 1-form
\begin{equation}
  E_{\vect{X}} := - i_{\vect{X}}\omega
\label{EXdef} \end{equation} is zero at a given point
$x\in\mathcal{M}$ if and only if $\vect{X}\at{x}=0$. As a
corollary, given a 1-form $\alpha$ on $\mathcal{M}$, there exists
one and only one vector field $\vect{X}$ such that $E_{\vect{X}}=
- i_{\vect{X}}\omega = \alpha$. Therefore,  eq. (\ref{EXdef})
defines a linear isomorphism from the tangent space
$T_x\mathcal{M}$ to the cotangent space $T^*_x\mathcal{M}$ at
every point $x$, and hence a linear isomorphism from the space of
vector fields $\mathcal{X(M)}$ to the space of differential
1-forms $\Omega^1(\mathcal{M})$. When the vector field $\vect{X}$
is as shown in
eq.~(\ref{Xdef}),
\begin{displaymath}
  E_{\vect{X}} = \dot{q}^{\,i}\,\dd p_i - \dot{p}_i\,\dd q^i.
\end{displaymath} Thus, the corresponding
Euler-Lagrange 1-form $E$ in eq.~(\ref{ELham}) becomes
\begin{equation}
  E = E_{\vect{X}} - \dd H.
\end{equation} As is known, for a Hamiltonian
function $H$ on $\mathcal{M}$, there exists uniquely a vector
field \begin{equation}
  \vect{X}_H := \frac{\partial H}{\partial p_i}\frac{\partial}{\partial q^i}
  - \frac{\partial H}{\partial q^i}\frac{\partial}{\partial p_i}
\end{equation} satisfying \begin{equation}
  E_{\vect{X}_H} = - i_{\vect{X}_H}\omega = \dd H.
\end{equation} Therefore the Euler-Lagrange
1-form
  $E = E_{\vect{X}} - E_{\vect{X}_H} = E_{\vect{X} - \vect{X}_H}$.
Since $\vect{X}$ can be an arbitrary vector field on $\mathcal{M}$
and so does $\vect{X}-\vect{X}_H$, it follows that
$\OmegaEL(\mathcal{M})$ is equal to $\Omega^1(\mathcal{M})$, and
that every 1-form on $\mathcal{M}$ is an Euler-Lagrange 1-form.
Thus, an immediate corollary is
$H_{\mathrm{EL}}(\mathcal{M},\omega)$ is equivalent to the first
de~Rham cohomology group
  $H^1_{\mathrm{dR}}(\mathcal{M})$.

For a symplectic manifold $(\mathcal{M},\omega)$, a vector field
$\vect{X}$ is called a
\deff{symplectic vector field} provided that
  $\dd E_{\vect{X}} = 0$.
A vector field $\vect{X}$ is called a
\deff{Hamiltonian vector field} provided that
  $E_{\vect{X}} = \dd H$
with some function $H$ on $\mathcal{M}$. \omits{Since the
symplectic form $\omega$ is closed, }The Lie derivative of
$\omega$ with respect to a vector field $\vect{X}$ reads
  $\Lied{X}\omega = \dd i_{\vect{X}}\omega = - \dd E_{\vect{X}}$.
  This
implies that a vector field $\vect{X}$ is symplectic if and only
if
  $\Lied{X}\omega = 0$.
A Hamiltonian vector field is, of course, a symplectic vector
field, but a symplectic vector field is not necessarily a
Hamiltonian vector field rather a local Hamiltonian vector field
\omits{(see below)}. These are the well-known facts in 
symplectic geometry \cite{lecture}.

In addition, the commutation bracket of two symplectic vectors
$\vect{X}$ and $\vect{Y}$ is a Hamiltonian vector field. In fact,
it is easy to
obtain \cite{lecture} that  %
\begin{displaymath}
  E_{[\vect{X,Y}]} = \dd\,\big(\omega(\vect{X,Y})\big),
  \qquad\textrm{ equivalently,}\qquad
  [\vect{X,Y}] = \vect{X}_{\omega(\vect{X,Y})}.
\end{displaymath} It implies that the linear
space \begin{displaymath}
  \XS(\mathcal{M},\omega) := \{\,\vect{X}\in\mathcal{X(M)}\,|\,\vect{X}
  \textrm{\ is\  symplectic}\,\}
\end{displaymath} is a Lie algebra with an ideal
\begin{displaymath}
  \XH(\mathcal{M},\omega) := \{\,\vect{X}\in\mathcal{X(M)}\,|\,\vect{X}
  \textrm{\ is\  Hamiltonian}\,\}.
\end{displaymath} This is due to
  $$[\XS(\mathcal{M},\omega),\XH(\mathcal{M},\omega)] \subseteq
   [\XS(\mathcal{M},\omega),\XS(\mathcal{M},\omega)] \subseteq
   \XH(\mathcal{M},\omega).$$
\omits{What we are interested in is, then, how about the quotient
Lie algebra
  $\XS(\mathcal{M},\omega)/\XH(\mathcal{M},\omega)$?}

It is obvious that
$\XS(\mathcal{M},\omega)/\XH(\mathcal{M},\omega)$ is an Abelian
Lie algebra. As we have stated, the linear map from
$\mathcal{X(M)}$ to $\Omega^1(\mathcal{M})$, sending $\vect{X}$ to
$E_{\vect{X}} = - i_{\vect{X}}\omega$, is a linear isomorphism.
The images of $\XS(\mathcal{M},\omega)$ and
$\XH(\mathcal{M},\omega)$ under this isomorphism are 
the spaces of closed 1-forms $Z^1(\mathcal{M})$ and the exact
1-forms $B^1(\mathcal{M})$ on $\mathcal{M}$, respectively. Hence
the linear isomorphism
  $E:\XS(\mathcal{M},\omega)\longrightarrow Z^1(\mathcal{M}),
  \vect{X}\longmapsto E_{\vect{X}}$
induces a linear isomorphism \begin{eqnarray*}
  \bar{E}: \XS(\mathcal{M},\omega)/\XH(\mathcal{M},\omega) & \longrightarrow &
  Z^1(\mathcal{M})/B^1(\mathcal{M}) \\
  \lbrack\vect{X}\rbrack & \longmapsto &
  [E_{\vect{X}}].
\end{eqnarray*} Namely, an isomorphism from the
quotient Lie algebra
$\XS(\mathcal{M},\omega)/\XH(\mathcal{M},\omega)$ to the first
de~Rham cohomology group $H^1_{\mathrm{dR}}(\mathcal{M})$, where
$[\vect{X}]$ is the equivalence class of the symplectic vector
field $\vect{X}$ in
$\XS(\mathcal{M},\omega)/\XH(\mathcal{M},\omega)$ and
$[E_{\vect{X}}]$ is the cohomology class of the closed 1-form
$E_{\vect{X}}$. All the above are summerized in the following
theorem:
\begin{thm}
  Under the commutation bracket, $\XS(\mathcal{M},\omega)$ is a Lie algebra
  with an ideal
$\XH(\mathcal{M},\omega)$. The quotient Lie algebra
$\XS(\mathcal{M},\omega)/\XH(\mathcal{M},\omega)$ is Abelian, and
linearly isomorphic to $H^1_{\mathrm{dR}}(\mathcal{M})$, hence to
the first Euler-Lagrange cohomology group
$H_{\mathrm{EL}}(\mathcal{M},\omega)$. \label{thm:thm4} \end{thm}

Therefore, if the symplectic manifold $(\mathcal{M},\omega)$ is
non-trivial such that $H^1_{\mathrm{dR}}(\mathcal{M})\neq 0$,
there exists a symplectic vector field that is not Hamiltonian.
Such a symplectic vector field can be locally written as
\begin{displaymath}
  \frac{\partial H}{\partial p_i}\frac{\partial}{\partial q^i}
  - \frac{\partial H}{\partial q^i}\frac{\partial}{\partial p_i}
\end{displaymath} with $H$ defined merely on a
proper open subset of $\mathcal{M}$. But there is not a globally
defined Hamiltonian function for it. It is in this sense that it
can be called a local Hamiltonian vector field \cite{Arnold}.

Note that although both $\XS(\mathcal{M}, \omega)$ and
$\XH(\mathcal{M},\omega)$ depend on the choice of the symplectic
structure, the quotient Lie algebra is, however, independent of
it. Therefore, we can always indicate, without specifying the
particular symplectic structure, how many linearly independent
local symplectic vector fields there are on $\mathcal{M}$. This is
the significance of the above theorem.

On the other hand, the above theorem also appears as an exact
sequence \cite{lecture}
\begin{displaymath}
  0 \longrightarrow \XH(\mathcal{M},\omega) \longrightarrow \XS(\mathcal{M},
  \omega) \longrightarrow H^1_{\mathrm{dR}}(\mathcal{M}) \longrightarrow
  0.
\end{displaymath}

\section{The Euler-Lagrange Cohomology  Groups on Symplectic  Manifolds}
\label{sectELsymp1}

Now we are ready to present the definition and study the
properties of a sequence of the Euler-Lagrange cohomology groups on a
$2n$-dimensional symplectic manifold $({\cal M},\omega)$.

Let $\Lambda^k(T^*_x \cal M)$ be the space of $k$-forms at a point
$x \in \cal M$, $\Lambda^k(\cal M)$ the corresponding fibre
bundle, $\Lambda^*_x(\cal M)$ the direct sum
$\bigoplus_{k=0}^{2n}\Lambda^k(T^*_x \cal M)$, $\Lambda^*(\cal M)$
the exterior bundle of $\cal M$, $\Omega^k(\cal M)$ the space of
differential $k$-forms, and $\Omega^*({\cal M})$ the exterior
algebra of differential forms.

\subsection{The $2k-1$st Euler-Lagrange
Cohomology Groups} \label{sectkth}

On $(\mathcal{M}^{2n},\omega)$, \omits{be a $2n$-dimensional
symplectic manifold. F}for each integer $1 \leqslant k \leqslant
n$ we can define two sets \begin{eqnarray}\label{Vs2k-1}
  \XS^{2k-1}(\mathcal{M},\omega) & := &
  \{\,\vect{X}\in\mathcal{X}(\mathcal{M})\,|\,\Lied{X}(\omega^k) = 0\,\},
\\\label{Vh2k-1}
  \XH^{2k-1}(\mathcal{M},\omega) & := &
  \{\,\vect{X}\in\mathcal{X}(\mathcal{M})\,|\,-i_{\vect{X}}(\omega^k) \textrm{\ is\ exact}
  \,\},
\end{eqnarray} which are obviously linear spaces
over $\mathbb{R}$. In the above,
\omits{$\vect{X}\in\mathcal{X}(\mathcal{M})$ is a vector field on
$M$,} $\omega^k$ is the wedge product of $k$-fold $\omega$. In
certain cases, we use the convention that $\omega^0 = 1$.
Obviously, for $k=1$, (\ref{Vs2k-1}) and (\ref{Vh2k-1}) give rise
to the symplectic vector fields and the Hamiltonian ones,
respectively:
\begin{equation}
  \XS^1(\mathcal{M},\omega) = \XS(\mathcal{M},\omega)
  \qquad \mathrm{and} \qquad
  \XH^1(\mathcal{M},\omega) = \XH(\mathcal{M},\omega).
\end{equation}

It can be found that
$\XS^{2n-1}(\mathcal{M},\omega)$ is the space of
volume-preserving vector fields. Since, for
arbitrary vector field $\vect{X}$, there is
\begin{equation}
  \Lied{X}(\omega^k) 
  = \dd i_{\vect{X}}(\omega^k),
\label{Liedk} \end{equation} a vector field
$\vect{X}$ belongs to $\XS(\mathcal{M},\omega)$
if and only if $ - i_{\vect{X}}(\omega^k)$ is
closed. Therefore, we obtain immediately
\begin{equation}
  \XH^{2k-1}(\mathcal{M},\omega) \subseteq
  \XS^{2k-1}(\mathcal{M},\omega),~~\forall k.
\end{equation} Since
$
  \Lied{X}(\omega^{k+1}) = \frac{k+1}{k}\,\Lied{X}(\omega^k)\wedge\omega
$ 
and
  $i_{\vect{X}}(\omega^{k+1})
   = \frac{k+1}{k}\,i_{\vect{X}}(\omega^k)\wedge\omega,$
it is also obvious that
\begin{eqnarray}
  & & \XS^1(\mathcal{M},\omega) \subseteq \ldots \subseteq
  \XS^{2k-1}(\mathcal{M},\omega) \subseteq \XS^{2k+1}(\mathcal{M},\omega)
  \subseteq \ldots \subseteq\XS^{2n-1}(\mathcal{M},\omega), \\
  & & \XH^1(\mathcal{M},\omega) \subseteq \ldots \subseteq
  \XH^{2k-1}(\mathcal{M},\omega) \subseteq \XH^{2k+1}(\mathcal{M},\omega)
  \subseteq \ldots \subseteq\XH^{2n-1}(\mathcal{M},\omega).
\end{eqnarray}
Similarly to the derivation of
  $[\XS(\mathcal{M},\omega), \XS(\mathcal{M},\omega)]
  \subseteq\XH(\mathcal{M},\omega)$,
We can verify that for arbitrary
  $\vect{X}$, $\vect{Y}\in \XS^{2k-1}(\mathcal{M},\omega)$
there is always
$\vect{[X,Y]}\in\XH^{2k-1}(\mathcal{M},\omega)$.
Namely, \begin{equation}
  [\XS^{2k-1}(\mathcal{M},\omega),\XS^{2k-1}(\mathcal{M},\omega)]
  \subseteq \XH^{2k-1}(\mathcal{M},\omega), ~~\forall k.
\end{equation} This indicates that
$\XH^{2k-1}(\mathcal{M},\omega)$ is an ideal of
$\XS^{2k-1}(\mathcal{M},\omega)$. Hence \deff{the $2k-1$st
Euler-Lagrange cohomology group (of degree $2k-1$)} can be defined
as a quotient Lie algebra \begin{equation}\label{defH_EL}
  \HEL^{2k-1}(\mathcal{M},\omega) :=
  \XS^{2k-1}(\mathcal{M},\omega)/\XH^{2k-1}(\mathcal{M},\omega),
\end{equation} which is Abelian for each $k$.
\omits{And it can be seen directly that
  $\HEL^1(M,\omega)\cong
  \HEL(M,\omega)$
due to Theorem \ref{thm:thm4}.}

On the other hand, for each $k$ ($1 \leqslant k
\leqslant n$),  the Euler-Lagrange $2k-1$ forms
$E_{\vect{X}}^{(2k-1)}$ as well as the kernel and
image spaces of them with respect to $\dd$ may be
introduced: \begin{eqnarray}
E_{\vect{X}}^{(2k-1)}(\cal M)&:=&-i_{\vect{X}}(\omega^k),
  ~~\vect{X}\in\mathcal{X}(\mathcal{M}, \omega);\\
  Z_{\rm EL}^{2k-1}(\mathcal{M})&:=&\{E_{\vect{X}}^{(2k-1)}|~\dd E_{\vect{X}}^{(2k-1)} = 0\};\\
  B_{\rm EL}^{2k-1}(\mathcal{M})&:=&\{E_{\vect{X}}^{(2k-1)}|E_{\vect{X}}^{(2k-1)}
  \textrm{ is exact }\}.
\end{eqnarray}
\omits{where $\omega^k$ is the wedge product of
$k$-fold $\omega$. We use the convention that
$\omega^0 = 1$.}
\deff{The $2k-1$st Euler-Lagrange cohomology
group} may also be equivalently defined as
\begin{equation}
  H^{2k-1}_{\mathrm{EL}}(\mathcal{M},\omega) :=Z_{\rm EL}^{2k-1}(\mathcal{M})
  /B_{\rm EL}^{2k-1}(\mathcal{M}).
\end{equation}
\omits{Then Lemma~\ref{lem:inject} implies that
this definition is equivalent to that in
eq.~(\ref{defH_EL}).}

\subsection{Some Operators}
\label{sect:Operators}

In order to investigate the properties of the
Euler-Lagrange cohomology groups, it is needed to
introduce some operators.

For a point $x \in \mathcal{M}$, the symplectic form
$\omega$\omits{, when restricted on the open set $U$,} can be
locally expressed as the well-known formula $\omega = \dd
p_i\wedge\dd q^i$ in the Darboux coordinates $(q,p)$.
Then let us introduce a well defined linear map \omits{
$i_{\frac{\partial}{\partial q^i}}i_{\frac{\partial}{\partial
p_i}}$} on $\Lambda^*_x(\mathcal{M})$: 
\begin{equation}
  \f := i_{\frac{\partial}{\partial q^i}}i_{\frac{\partial}{\partial p_i}}.
\label{eq:f} \end{equation} Note that $\f = 0$ when acting on
$\Lambda^1(T^*_x \mathcal{M})$ or $\Lambda^0(T^*_x
\mathcal{M})$.\omits{, and that the operator $\f$ is well
defined.} And a map $\f$ can be defined on the exterior bundle
$\Lambda^*(\mathcal{M})$ point by point. Further, a linear
homomorphism, denoted also by $\f$, can be obtained on
$\Omega^*(\mathcal{M})$. Especially, we have the identity
\begin{equation}
  \f\omega = n.
\label{pomega}
\end{equation}

Another two operators
\begin{equation}
  \e: \Lambda^*_x(\mathcal{M}) \longrightarrow \Lambda^*_x(\mathcal{M}), \quad
  \alpha \longmapsto \e\alpha = \alpha\wedge\omega
\label{eq:e}
\end{equation}
and
\begin{equation}
  \h: \Lambda^k(T^*_x \mathcal{M})\longrightarrow \Lambda^k(T^*_x \mathcal{M}), \quad
  \alpha \longmapsto \h\alpha = (k-n)\,\alpha
\label{eq:h} \end{equation} can also be defined
at each $x\in \mathcal{M}$. \begin{lem}
  The operators $\e$, $\f$ and $\h$ on $\Lambda^*_x(\mathcal{M})$ satisfy
  \begin{equation}
    [\h,\e] = 2\,\e, \qquad
    [\h,\f] = -2\,\f, \qquad
    [\e,\f] = \h, \qquad \forall\,x \,\in\, \cal M .
  \label{eq:efh}
  \end{equation}
\label{lem:efh}
\end{lem}
\begin{proof}
These relations can be verified directly. Here is a trickier proof.

First we define some ``fermionic" operators on
$\Lambda^*_x(\mathcal{M})$
\begin{equation}
\begin{array}{lll}
  \psi_i := i_{\frac{\partial}{\partial q^i}}, & \qquad &
  \psi^i := i_{\frac{\partial}{\partial p_i}}, \\
  \chi_i : \alpha \longmapsto \dd p_i\wedge\alpha, & &
  \chi^i : \alpha \longmapsto \dd q^i\wedge\alpha.
\end{array} \label{eq:fermionic} \end{equation}
For these operators, it is easy to verify that the non-vanishing
anti-commutators  are \begin{equation}
  \{\psi_i,\chi^j\} = \delta^j_i, \qquad
  \{\psi^i,\chi_j\} = \delta^i_j.
\end{equation} \omits{This can be verified easily.} Given
an integer $0 \leqslant k \leqslant 2n$, we can check that, for
any $\alpha\in\Lambda^k(T^*_x \mathcal{M})$, \begin{equation}
  (\chi_i\,\psi^i + \chi^i\,\psi_i)\alpha = k\,\alpha.
\end{equation}
Therefore,
\begin{equation}
  \h = \chi_i\,\psi^i + \chi^i\,\psi_i - n.
\end{equation}
According to the definitions,
\begin{equation}
  \e = \chi_i\,\chi^i, \qquad \f = \psi_i\,\psi^i.
\end{equation}
Then the relations in eqs.~(\ref{eq:efh}) can be obtained when $\e$,
$\f$ and $\h$ are viewed as bosonic operators.
\end{proof}

\comments{
\begin{corol}
  Let $x \in M$ be an arbitrary point and $0 \leqslant k \leqslant 2n$.
  If $\alpha \in \Lambda^k(T^*_x M)$ satisfies $\alpha\wedge\omega = 0$, then
  $(\f\alpha)\wedge\omega = - (n-k)\, \alpha.$
\label{lem:kernelomega}
\end{corol}
}

For a point $x \in \mathcal{M}$ and $\alpha \in \Lambda^k(T^*_x
\mathcal{M})$, $(0 \leqslant k \leqslant 2n)$, the following
formulae can be derived recursively: \begin{equation}
  [\e^k,\f] = k\,\e^{k-1}(\h + k - 1), \qquad
  [\e,\f^k] = k\,\f^{k-1}(\h - k + 1),
\label{eq:efl}
\end{equation}
where $k$ is an arbitrary positive integer. Then there is the
lemma:

\begin{lem}
  Let $\alpha$ be a 2-form. If $\e^k \alpha=0$ for some $k<n-1$, then
$\alpha=0$.
\label{lem:injective}
\end{lem}
\begin{proof}
  Applying both sides of the first eqn in (\ref{eq:efl}) on $\alpha$, we have
  \begin{equation}
    \e^k \f \alpha = k \, (k-n+1)\, \e^{k-1} \alpha.
  \label{eq:wsy}
  \end{equation}
  Since $\f\alpha$ is a number at each point, the left hand side is
  $(\f\alpha)\, \omega^k$. Applying $\e$ on both sides, we get
  $$ (\f\alpha)\, \omega^{k+1}= k(k-n+1)\, \e^k \alpha=0. $$
  Since $k+1<n$, we have $\f\alpha=0$. Now formula (\ref{eq:wsy}) becomes
  $$  k \, (k-n+1)\, \e^{k-1} \alpha=0. $$
  Since $k<n-1$, we get $\e^{k-1} \alpha=0.$ Therefore, the value of $k$ can be
  reduced by 1, and further it can be eventually reduced to 0.
\end{proof}

The above lemma implies that the map sending
$\alpha\in\Lambda^2(T^*_x \mathcal{M})$ to
$\alpha\wedge\omega^{n-2}\in\Lambda^{2n-2}(T^*_x \mathcal{M})$ is
an isomorphism.

\begin{lem}
  Let $x\in \mathcal{M}$ be an arbitrary point and $\vect{X}\in T_x \mathcal{M}$. Then, for each
$1 \leqslant k \leqslant n$, $i_{\vect{X}}(\omega^k) = 0$ if and
only if $\vect{X} = 0$. \label{lem:inject}
\end{lem}
\begin{proof}
We need only to prove that $i_{\vect{X}}(\omega^k) = 0$ implies
$\vect{X} = 0$. We assume that there is a nonzero vector
$\vect{X}\in T_x \mathcal{M}$ satisfying $i_{\vect{X}}(\omega^k) =
0$ for some $k$.

Immediately we have $i_{\vect{X}}(\omega^n) = 0$. Since $\vect{X}$
is nonzero, a basis $\{\vect{X}_1, \ldots, \vect{X}_{2n}\}$ can
always be obtained where $\vect{X}_1 = \vect{X}$. Consequently, it
follows\omits{ According to these, we can say} that
$\omega^n(\vect{X}_1,\ldots,\vect{X}_{2n}) =
(i_{\vect{X}}(\omega^n))(\vect{X}_2,\ldots,\vect{X}_{2n}) = 0$.
However, this contradicts with the fact that $\omega$ is
non-degenerate. Therefore $\vect{X}$ has to be zero.
\end{proof}

\subsection{The Spaces
$\XS^{2k-1}(\mathcal{M},\omega)$ and
  $H_{{\rm EL}}^{2n-1}(\mathcal{M},\omega)$}

In the subsection \ref{sectkth}, we have indicated that
$\XS(\mathcal{M},\omega) = \XS^1(\mathcal{M},\omega) \subseteq
\XS^{2k-1}(\mathcal{M},\omega)$ for each possible $k$. The
following theorem tells us much more.
\begin{thm}
  Let $(\mathcal{M},\omega)$ be a $2n$-dimensional symplectic manifold with
$n \geqslant 2$. Then, for each $ k \in \{1, 2, \ldots,  n - 1 \}$,
  \begin{equation}
  \XS^{2k-1}(\mathcal{M},\omega) = \XS(\mathcal{M},\omega).
  \end{equation}
\label{thm:XS}
\end{thm}
\begin{proof} We need only to prove that
$\XS^{2k-1}(\mathcal{M},\omega) \subseteq \XS(\mathcal{M},\omega)$
for each $k\in \{1, 2, \dots, n-1\}$.

In fact, for any
$\vect{X}\in\XS^{2k-1}(\mathcal{M},\omega)$,
\begin{displaymath}
  \Lied{X}(\omega^k) = k\,(\Lied{X}\omega)\wedge\omega^{k-1} = 0.
\end{displaymath} Since $0 \leqslant k-1
\leqslant n-2$ while $\Lied{X}\omega$ is a 2-form, we can use
Lemma \ref{lem:injective} pointwisely. This yields $\Lied{X}\omega
=0$. Thus, $\vect{X}\in\XS(\mathcal{M},\omega)$. This proves
$\XS^{2k-1}(\mathcal{M},\omega)\subseteq\XS(\mathcal{M},\omega)$
when $1 \leqslant k \leqslant n-1$.
\end{proof}

As was implied by Lemma \ref{lem:inject}, the map
$\mathcal{X}(\mathcal{M})\longrightarrow\Omega^{2n-1}(\mathcal{M}),
\vect{X}\longmapsto i_{\vect{X}}(\omega^n)$ is a linear
isomorphism. From $\Lied{X}(\omega^n) = \dd
i_{\vect{X}}(\omega^n)$, it follows that
$\XS^{2n-1}(\mathcal{M},\omega)$ is isomorphic to
$Z^{2n-1}(\mathcal{M})$, the space of closed $(2n-1)$-forms. Lemma
\ref{lem:inject} also implies that
$\XH^{2n-1}(\mathcal{M},\omega)$ is isomorphic to
$B^{2n-1}(\mathcal{M})$, the space of exact $(2n-1)$-forms. These
can be summarized as in the following theorem: \begin{thm} The
linear map $\nu_n:
\mathcal{X}(\mathcal{M})\longrightarrow\Omega^{2n-1}(\mathcal{M}),
\vect{X} \longmapsto i_{\vect{X}}(\omega^n)$ is an isomorphism.
Under this isomorphism, $\XS^{2n-1}(\mathcal{M},\omega)$ and
$\XH^{2n-1}(\mathcal{M},\omega)$ are isomorphic to
$Z^{2n-1}(\mathcal{M})$ and $B^{2n-1}(\mathcal{M})$, respectively.
\label{thm:thm6} \end{thm}

\begin{corol}
  The $(2n-1)$st Euler-Lagrange cohomology group
$H^{2n-1}_{\mathrm{EL}}(\mathcal{M},\omega)$ is linearly
isomorphic to $H^{2n-1}_{\mathrm{dR}}(\mathcal{M})$, the
$(2n-1)$st de~Rham cohomology group. \end{corol}

When $\mathcal{M}$ is closed,
$\HEL^{2n-1}(\mathcal{M},\omega)$ is linearly
isomorphic to the dual space of
$\HEL^1(\mathcal{M},\omega)$, because
  $\HdR^k(\mathcal{M}) \cong (\HdR^{2n-k}(\mathcal{M}))^*$
for such kind of manifolds. If $\mathcal{M}$ is not compact, this
relation cannot be assured.

\subsection{The Other Euler-Lagrange Cohomology Groups}
\label{sectRelation}

Although the first and the last Euler-Lagrange cohomology groups
can be identified with the corresponding de~Rham cohomology
groups, respectively, it is still valuable to know whether the
other Euler-Lagrange cohomology groups are nontrivial and
different from corresponding de Rahm cohomology groups in
general.\omits{, are still unknown.}

In this subsection we will enumerate some examples and properties
to seeing about this problem. We shall  point out that, for the
torus $T^{2n}$ with the standard symplectic structure $\omega$ and
$n \geqslant 3$, $\HEL^{2k-1}(T^{2n}, \omega)$ is not isomorphic
to $\HdR^{2k-1}(T^{2n})$ whenever $1 < k < n$ (see,
Corollary~\ref{cor:T2n}). In addition, we shall prove that there
is a $6$-dimensional symplectic manifold $({\cal M},\omega)$ for
which the Euler-Lagrange cohomology group $\HEL^3
(\mathcal{M},\omega)$ is not isomorphic to $\HdR^1(\mathcal{M})$
(see, Theorem~\ref{thm:nilm}). Therefore, these indicate that the
Euler-Lagrange cohomology groups other than the first and the last
ones are some new features of certain given symplectic manifolds.

Let $L_\alpha$ be the homomorphism defined by the cup product with
a cohomology class $[\alpha]$, where
 $\alpha$ is a representative. From the
definition, there is an injective homomorphism of vector spaces
$$\pi_{2k-1}: H^{2k-1}_{\mathrm{EL}}(\mathcal{M},\omega)
\longrightarrow
  H^{2k-1}_{\mathrm{dR}}(\mathcal{M}) $$
for each $ k \in \{ 1, 2, \ldots, n \}$ such that the following diagram is
commutative:
\begin{equation}
\begin{CD}
  H^1_{\mathrm{EL}}(\mathcal{M},\omega) @> >> H^3_{\mathrm{EL}}(\mathcal{M},\omega) @> >> \cdots
  @> >> H^{2n-3}_{\mathrm{EL}}(\mathcal{M},\omega)
\\
  @V \pi_1 VV   @V \pi_3 VV    @V \cdots VV    @V \pi_{2n-3} VV
\\
  H^1_{\mathrm{dR}}(\mathcal{M}) @> L_{\omega} >> H^3_{\mathrm{dR}}(\mathcal{M}) @> L_{\omega} >>
                \cdots @> L_{\omega} >> H^{2n-3}_{\mathrm{dR}}(\mathcal{M}).
\end{CD} \label{eq:CD} \end{equation} In fact,
for an equivalence class $[\vect{X}]_{2k-1} \in
H^{2k-1}_{\mathrm{EL}}(\mathcal{M},\omega)$ $(1 \leqslant k
\leqslant n)$ with $\vect{X} \in \XS^{2k-1}(\mathcal{M},\omega)$
an arbitrary representative, $ -
\frac{1}{k}\,i_{\vect{X}}(\omega^k)$ is a closed $(2k - 1)$-form.
Thus the cohomology class of this form can be defined to be
$\pi_{2k-1}([\vect{X}]_{2k-1})$ and it is easy to verify that this
definition is well defined. As for the horizontal maps in the
first row of the above diagram, they are induced by the identity
map on $\XS^{2k-1}(M,\omega)=\XS^1(\mathcal{M},\omega)$ where $k
\in \{1,2,\ldots,n-1\}$. For example, if $[\vect{X}]_{2k-1}$ is an
equivalence class in $H^{2k-1}_{\mathrm{EL}}(\mathcal{M},\omega)$
$( k=1, 2, \ldots, n-1,\ n > 1 )$ where
$\vect{X}\in\XS^{2k-1}(\mathcal{M},\omega)$ is an arbitrary
representative, then $[\vect{X}]_{2k-1}$ is mapped to be an
equivalence class $[\vect{X}]_{2k+1}$ in
$H^{2k+1}_{\mathrm{EL}}(\mathcal{M},\omega)$. It is also easy to
check that this is a well defined homomorphism.

Since $\pi_1$ is an isomorphism and the horizontal homomorphisms
in the first row are all onto, it follows that \begin{thm} For $2
\leqslant k \leqslant n-1$, $\pi_{2k-1}$ is onto if and only if
$L_{\omega}^{k-1}=L_{\omega^{k-1}}$ from
$H^1_{\mathrm{dR}}(\mathcal{M})$ to
$H^{2k-1}_{\mathrm{dR}}(\mathcal{M})$ is onto, and
$L_{\omega^{k-1}} : \HdR^1(\mathcal{M}) \longrightarrow
\HdR^{2k-1}(\mathcal{M})$ is injective if and only if the
homomorphism from $\HEL^1(\mathcal{M},\omega)$ to
$\HEL^{2k-1}(\mathcal{M},\omega)$ is injective. \label{T:dec}
\end{thm}

\begin{corol} For $n \geqslant 3$, let
$\mathcal{M}$ be the torus $T^{2n}$ with the standard symplectic
structure $\omega$. Then, for $1 < k < n$,
$\HEL^{2k-1}(\mathcal{M},\omega) \neq \HdR^{2k-1}(\cal M)$.
\label{cor:T2n}
\end{corol}
\begin{proof} As is well known, the de~Rham cohomology groups
of $T^{2n}$ satisfy \begin{displaymath}
  \dim \HdR^k(T^{2n}) = {{2n}\choose{k}}
\end{displaymath}
for each $0 \leqslant k \leqslant 2n$. Therefore, we have
$\dim\HdR^{2k-1}(T^{2n}) > 2n$ for each $1 < k < n$.
On the other hand, due to the fact that the maps in the first row of the
diagram (\ref{eq:CD}) are surjective, we have
$\dim\HEL^{2k-1}(T^{2n},\omega) \leqslant 2n$ for each $1 < k < n$.
So, $\dim\HdR^{2k-1}(T^{2n}) >\dim\HEL^{2k-1}(T^{2n},\omega)$.
\end{proof}

In what follows, we will show further that there are some
symplectic manifolds for which $\HEL^k \neq \HEL^1$.

Recall that on an $n$-dimensional Lie group $G$ there exists
a basis that consists of $n$ left-invariant vector fields
$\vect{X}_1$, \ldots, $\vect{X}_n$. They form the Lie algebra
$\mathfrak{g}$ of $G$.\omits{ can be chosen to be the frame.} Let
$[\vect{X}_i, \vect{X}_j] =- c_{ij}^k\,\vect{X}_k$ with  the
structural constants $c_{ij}^k$ of $\mathfrak{g}$. Let $\{\theta^k\}$
be the left-invariant dual basis, i.e.
\begin{equation}
  \dd\theta^k =\frac{1}{2}\,c_{ij}^k\,\theta^i \wedge\theta^j,
  \quad
k=1,\ldots, n.
\end{equation}

\omits{Let $\mathfrak{g}$ be the Lie algebra of $G$.} $G$ is
called a
\deff{nilpotent} Lie group if $\mathfrak{g}$ is nilpotent. A {\it nilmanifold}
is defined to be a closed manifold $M$ of the form $G/\Gamma$
where $G$ is a simply connected nilpotent group and $\Gamma$ is a
discrete subgroup of $G$. It is well known that $\Gamma$
determines $G$ and is determined by $G$ uniquely up to isomorphism
(provided that $\Gamma$ exists) \cite{M,R}.

There are three important facts for the compact nilmanifolds
\cite{TO}:
\begin{enumerate} \item Let $\mathfrak{g}$ be a nilpotent Lie algebra with
structural constants $c_{ij}^k$ with respect to some basis, and
let $\{\theta^1,\ldots ,\theta^n\}$ be the dual basis of
$\mathfrak{g}^{\ast}$. Then in the Chevalley--Eilenberg complex
$(\Lambda^{\ast}\mathfrak{g}^{\ast},\dd)$ we have
\begin{equation}\label{cte}
\dd\theta^k=\sum_{1\leqslant i<j<k} c_{ij}^k \,\theta^i\wedge
\theta^j, \quad k=1,\ldots, n.
\end{equation}
\item Let $\mathfrak{g}$ be the Lie algebra of a simply connected nilpotent
Lie group $G$. Then, by Malcev's theorem \cite{M}, $G$ admits a lattice if
and only if $\mathfrak{g}$ admits a basis such that all the structural
constants are rational.
\item By Nomizu's theorem, the Chevalley-Eilenberg complex
$(\Lambda^{\ast}\mathfrak{g}^{\ast},\dd)$ of $\mathfrak{g}$ is quasi-isomorphic
to the de~Rham complex of $G/\Gamma$. In particular,
\begin{equation}\label{nom}
H^{\ast}(G/\Gamma)\cong H^{\ast}(\Lambda^{\ast}\mathfrak{g}^*,\dd)
\end{equation}
and any cohomology class $[\alpha]\in H^k(G/\Gamma)$ contains a
homogeneous representative $\alpha$. Here we call the form
$\alpha$ homogeneous if the pullback of $\alpha$ to $G$ is
left-invariant.
\end{enumerate}

These results allow us to compute cohomology invariants of
nilmanifolds in terms of the Lie algebra $\mathfrak{g}$, and this simplifies
the calculations.

\begin{thm} There exists a $6$-dimensional
symplectic nilmanifold $(\mathcal{M},\omega)$ such that
$$\HEL^{3}(\mathcal{M},\omega)\neq \HEL^1({\cal M},\omega).$$
\label{thm:nilm}
\end{thm}
\begin{proof} To define the manifold $\mathcal{M}$, it
suffices to give the Lie algebra. $\mathfrak{g}$ is a
$6$-dimensional Lie algebra generated by the generators
$\vect{X}_1,\ldots,\vect{X}_6$ with Lie bracket given by
$$[\vect{X}_i,\vect{X}_j]=-\sum_{1\leqslant i<j<k} c_{ij}^k
\vect{X}_k.$$ This Lie algebra gives a unique nilmanifold
$\mathcal{M}$ by the above information on nilmanifolds. The
Chevalley-Eilenberg complex
$(\Lambda^{\ast}\mathfrak{g}^{\ast},\dd)$ of $\mathfrak{g}$ which
calculates the de~Rham cohomology of $\mathcal{M}$ is as follows.

Let $A=\Lambda^*(\theta^1,\ldots,\theta^6)$ with  the 1-forms
$\theta^i, ~1\leqslant i \leqslant 6$. Their differentials
are given by the following formulae:
\begin{displaymath}
\begin{array}{lcl}
  \dd \theta^4=\theta^1\wedge \theta^2 , & &
  \dd \theta^5=\theta^1\wedge \theta^4 -\theta^2\wedge \theta^3, \\
  \dd \theta^6=\theta^1\wedge \theta^5 +\theta^3\wedge \theta^4. & &
\end{array}
\end{displaymath}
Furthermore, the symplectic form
$\omega$ on $\mathcal{M}$ is induced by $F=\theta^1\wedge \theta^6
+\theta^2\wedge \theta^4 +\theta^3\wedge \theta^5$ in $A$.
$\omega$ is a symplectic form since $F\wedge F\wedge F$ is
nontrivial by an easy calculation.

It is not difficult to show
$$\HdR^1(M) = \mathbb{R}^3 = \Span\{[\theta^1],[\theta^2],[\theta^3]\}$$
and $$\HdR^2(M)=\mathbb{R}^4 = \Span\{[\theta^1\wedge \theta^3],
[\theta^1\wedge \theta^4], [\theta^2\wedge \theta^4], [F]\}$$

To prove the theorem, it follows from Theorem~\ref{T:dec} that it
is sufficient to prove that $\omega \wedge \theta^1$ is
cohomologically trivial. This follows by the following equation
that can be checked easily and thus completes the proof:
$$F \wedge \theta^1=\theta^1\wedge \theta^2\wedge \theta^4
+\theta^1\wedge \theta^3\wedge \theta^5
=\dd(\theta^2\wedge \theta^5 +\theta^3\wedge \theta^6).$$
\end{proof}


\subsection{The Euler-Lagrange Cohomology and The Harmonic Cohomology}

On a given symplectic manifold $(\mathcal{M},\omega)$, there also
exists the harmonic cohomology in addition to the de Rahm
cohomology.  In this subsection we explore the relation between
the Euler-Lagrange cohomology and the harmonic cohomology on
$(\mathcal{M},\omega)$ and show that they are different from each
other in general.

Given a smooth symplectic  manifold
$(\mathcal{M},\omega)$, let
$\Omega^k(\mathcal{M})$ be the space of all
$k$-forms on $\mathcal{M}^{2n}$. The $*$-operator
$$  *: \Omega^k(\mathcal{M}) \to
\Omega^{2n-k}(\mathcal{M})$$ can be introduced \cite{Y} in analog
with the $*$-operator on a Riemannian manifold. Define
$$\delta: \Omega^k(M) \to \Omega^{k-1}(M), \quad
\delta(\alpha)=(-1)^{k+1}*\dd(*\alpha). $$ It turns out to be that
$\delta=[i(\Pi),\dd]$ (see \cite{Y,Br}), where $i(\Pi)$ is, in
fact, the operator $\f$ introduced before.

{\bf Remark 1:} The operator $\delta=-*\dd*$ was also considered
by Libermann (see \cite{LM}). Koszul \cite{K} introduced the
operator $\delta=[\dd,i(\Pi)]$ for Poisson manifolds. Brylinski
\cite{Br} proved that these operators coincide.

{\bf Definition:} \rm A form $\alpha$ on a symplectic manifold
$(\mathcal{M}, \omega)$ is called {\it symplectically harmonic} if
$d\alpha =0=\delta \alpha$. 

We denote by $\Omega^k_{\mathrm{hr}}(M)$ the
linear space of symplectically harmonic
$k$-forms.  Unlike the Hodge theory, there are
non-zero exact symplectically harmonic forms.
Now, following Brylinski \cite{Br}, we define
symplectically harmonic cohomology
$H^*_{\mathrm{hr}}(\mathcal{M},\omega)$ by
setting $$
H^k_{\mathrm{hr}}(\mathcal{M},\omega)=\Omega^k_{\mathrm{hr}}(\mathcal{M})/(\im
(\dd) \cap
 \Omega^k_{\mathrm{hr}}(\mathcal{M})).$$
Therfore, $H^k_{\mathrm{hr}}(\mathcal{M},\omega) \subset
\HdR^k(\mathcal{M})$.

We would like to know if the symplectically harmonic cohomology
and the Euler-Lagrange cohomology are isomorphic to each other.
The following result answers this question.
\begin{thm} Let $\mathcal{M}$ be the $2n$
dimensional torus $T^{2n}$ with standard
symplectic structure. Then
$H^{2k-1}_{\mathrm{EL}}(\mathcal{M},\omega)$ and
$H^{2k-1}_{\mathrm{hr}}(\mathcal{M},\omega)$ are
not the same for $1<k<n$. \end{thm}

This is because in this case the symplectically harmonic cohomology
are the same as the de~Rham cohomology and now the result follows from
Corollary~\ref{cor:T2n}.

\section{The Relative Euler-Lagrange Cohomology}
\label{sect:RelativeEL}

Let us now propose a definition of relative
Euler-Lagrange cohomology that is the combination
of the above definition of Euler-Lagrange
cohomology and the usual definition of relative
de~Rham cohomology.

Let $\mathcal{M}$ be a $2n$-symplectic manifold and $i:
\mathcal{N} \longrightarrow \mathcal{M}$ be an embedded
sub\-manifold. Recall that the usual relative de~Rham forms are
defined by
  $$\Omega^k(i)=\Omega^k(\mathcal{M}) \oplus \Omega^{k-1}(\mathcal{N})$$
where $\Omega^{k-1}(\mathcal{N})$ is the group of
$(k-1)$-forms on $\mathcal{N}$. The differential
is given by \begin{equation}
  \dd(\theta_1,\theta_2)=(\dd\theta_1,i^*\theta_1 - \dd\theta_2).
\end{equation}
{\bf Definition:} Define
  $$\OmegaEL^{2k-1}(i)=\OmegaEL^{2k-1}(\mathcal{M})\oplus\Omega^{2k-2}(\mathcal{N})$$ where
  $\OmegaEL^{2k-1}(\mathcal{M})=\{i_{\vect{X}}(\omega ^k)\,|\, \vect{X}
  \in \mathcal{X}(\mathcal{M})\}$.
The \deff{relative Euler-Lagrange cohomology} will be defined as
$$\HEL^{2k-1}(i)=\frac{\{(\theta_1 , \theta_2)\in \OmegaEL^k(i) \,|\,
\dd(\theta_1 , \theta_2)=0 \}}{\{(\theta_1 ,
\theta_2)\,|\, (\theta_1 , \theta_2)=\dd(\theta'_1 , \theta'_2)\}}.$$

Let us consider an example for the relative
Euler-Lagrange cohomology. Let
$\mathcal{M}=\R^{2n}$, $\mathcal{N}=T^n$ and $i:
T^n \longrightarrow \R^{2n}$ be the inclusion.
\begin{proposition}
 $\HEL^{2k-1}(i)=H_{\rm d R}^{2k-2}(T^n).$
\end{proposition}
\begin{proof} There is the obvious linear map $\Omega^{2k-2}(T^n)
\longrightarrow \OmegaEL^{2k-1}(i)$ given by $\theta \longmapsto
(0,\theta)$ . When $\theta = \dd\alpha\in\Omega^{2k-2}(T^{2n})$ is
exact, $(0,\theta) = \dd(0,-\alpha)\in\OmegaEL^{2k-1}(i)$ is also
exact. Therefore a linear map $f:
\HdR^{2k-2}(T^{2n})\longrightarrow\HEL^{2k-1}(i),$ $[\theta]
\longmapsto [(0,\theta)]$ can be induced.

This map $f$ is an injection. In fact, if
$(0,\theta)=\dd(\alpha_1,\alpha_2)$, then $\dd \alpha_1=0$,
$\theta=i^*(\alpha_1)-\dd \alpha_2.$ Thus $\theta$ is exact by the
fact that any closed form on $\R^{2n}$ is exact.

The map $f$ is also an epimorphism: For any closed
$(\theta_1,\theta_2)$, it is in the same cohomology class of the
element $(0, \theta_2-i^*(\alpha_1)+\dd \alpha_2) =
(\theta_1,\theta_2) -\dd(\alpha_1, \alpha_2)$, where $\theta_1=\dd
\alpha_1$ and $\alpha_2$ is any form on $T^n$. Obviously,
$\theta_2-i^*(\alpha_1)+\dd \alpha_2$ is closed and
$f([\theta_2-i^*(\alpha_1)+\dd \alpha_2]) =
[(\theta_1,\theta_2)].$
\end{proof}

\vskip 2mm 
{\bf Remark 2:} Although it is not verified yet, the following
statement, if true, will not be a surprise : There exists a
symplectic manifold $\cal M$ and its submanifold $i:\mathcal{N}
\longrightarrow \mathcal{M}$ for which the (relative)
Euler-Lagrange cohomology is not the corresponding (relative)
de~Rham cohomology.

\vskip 2mm 
{\bf Remark 3:} For the definition of $\HEL^{2k-1}(i)$, it is also
possible to require that $(\theta'_1,\theta'_2)$ belong to
$\Omega^{2k-1}(i) =
\Omega^{2k-1}(\mathcal{M})\oplus\Omega^{2k-2}(\mathcal{N})$ rather
than $\OmegaEL^{2k-1}(i)$. The remaining explanations are similar
in principle.

\section{The General Volume-Preserving Hamiltonian-like Equations}

In this section, we present the general form for the equations of
the volume-preserving systems from the cohomological point of view
on a symplectic manifold $({\cal M}, \omega)$ as the phase space
of a kind of mechanical systems.

As was mentioned before, the canonical equations of a Hamiltonian
system belong to the image of the first Euler-Lagrange cohomology
group since the null Euler-Lagrange 1-form, which leads to the
canonical equations, belongs to the image. In this section, we
consider the highest Euler-Lagrange cohomology group and to show
its image may lead to the general volume-preserving equations on
symplectic manifolds and the ordinary canonical equations are
their special cases\cite{zgw, zgp}. Thus, this generalizes the
Hamiltonian systems.

\subsection{The Derivation of the Equations}

Theorem \ref{thm:thm6} indicates that there exists the 1-1 and
onto map $\nu_n$ between $\XH^{2n-1}(\mathcal{M},\omega)$ and
$B^{2n-1}(M)$. In addition, there also exists the following
operator:
  $$\star:\Omega^2(\mathcal{M})\longrightarrow\Omega^{2n-2}(\mathcal{M}),\qquad
  \alpha \longmapsto
\alpha\wedge\omega^{n-2}.$$ Since it is an isomorphism, there is a
unique linear map
$\phi:\Omega^2(\mathcal{M})\longrightarrow\XH^{2n-1}(\mathcal{M},\omega)$
making the diagram \begin{equation} \begin{CD}
  \Omega^2(\mathcal{M}) @>\star>> \Omega^{2n-2}(\mathcal{M}) \\
  @V\phi VV                        @V\dd VV\\
  \XH^{2n-1}(\mathcal{M},\omega) @>\nu_n>> B^{2n-1}(\mathcal{M})
\end{CD} \end{equation} commutative. Explicitly,
for an arbitrary
$\alpha\in\Omega^2(\mathcal{M})$, the
corresponding vector field in
$\XH^{2n-1}(\mathcal{M},\omega)$ is
\begin{equation}\label{phialpha}
  \phi\alpha = \nu_n^{-1}\dd\star\alpha.
\end{equation}

Now suppose that
\begin{equation}
  \alpha = \frac{1}{2}\,A_{ij}\,\dd q^i\wedge\dd q^j
  + A^i_j\,\dd p_i\wedge\dd q^j + \frac{1}{2}\,A^{ij}\,\dd p_i\wedge\dd p_j.
\end{equation}
Since $\dd\star\alpha =
(\dd\alpha)\wedge\omega^{n-2}$, we can obtain
that
\begin{eqnarray*}
  \dd\star\alpha & = & \frac{1}{2}\,\frac{\partial A_{ij}}{\partial q^k}\,
  \dd q^i\wedge\dd q^j\wedge\dd q^k\wedge\omega^{n-2}
  + \frac{1}{2}\,\frac{A^{ij}}{\partial p_k}\,
    \dd p_i\wedge\dd p_j\wedge\dd p_k\wedge\omega^{n-2}
\\
  & & + \bigg(\frac{1}{2}\,\frac{\partial A_{jk}}{\partial p_i}
    + \frac{\partial A^i_j}{\partial q^k}\bigg)\,
    \dd p_i\wedge\dd q^j\wedge\dd q^k\wedge\omega^{n-2}
 \\
 & & + \bigg( \frac{1}{2}\,\frac{\partial A^{ij}}{\partial q^k}
    -\frac{\partial A^i_k}{\partial p_j} \bigg)\,
    \dd p_i\wedge\dd p_j\wedge\dd q^k\wedge\omega^{n-2}
\\
  & = &  \bigg(\frac{1}{2}\,\frac{\partial A_{jk}}{\partial p_i}
    + \frac{\partial A^i_j}{\partial q^k}\bigg)\,
    \dd p_i\wedge\dd q^j\wedge\dd q^k\wedge\omega^{n-2}
 \\
 & & + \bigg( \frac{1}{2}\,\frac{\partial A^{ij}}{\partial q^k}
    -\frac{\partial A^i_k}{\partial p_j} \bigg)\,
    \dd p_i\wedge\dd p_j\wedge\dd q^k\wedge\omega^{n-2}.
\end{eqnarray*}

It is easy to see that, for arbitrary $i, j, k = 1, \ldots, n$,
\begin{equation}
  \dd p_i\wedge\dd q^k\wedge\dd p_j\wedge\dd q^l\wedge\omega^{n-2}
  = \frac{\delta^{kl}_{ij}}{n(n-1)}\,\omega^n ,
\end{equation}
where $\delta^{kl}_{ij} := \delta^k_i\,\delta^l_j -
\delta^l_i\,\delta^k_j$. Using $\frac{\partial}{\partial q^l}$ and
$\frac{\partial}{\partial p_j}$ to contract both sides of the
above equation and summing $l$ and $j$ over 1 to $n$, we can
obtain, respectively
\begin{eqnarray}
  \dd p_i\wedge\dd p_j\wedge\dd q^k \wedge\omega^{n-2}
  &=& \frac{\delta^k_j}{n-1}\,\dd p_i \wedge\omega^{n-1}
  - \frac{\delta^k_i}{n-1}\,\dd p_j \wedge\omega^{n-1},\\
  \dd p_i\wedge\dd q^j\wedge\dd q^k\wedge\omega^{n-2}
  &=& \frac{\delta^j_i}{n-1}\,\dd q^k\wedge\omega^{n-1}
  - \frac{\delta^k_i}{n-1}\,\dd q^j\wedge\omega^{n-1}.
\end{eqnarray}
By virtue of these two equations, we can write $\dd\star\alpha$ as
\begin{eqnarray}
  \dd\star\alpha & = & \frac{1}{n-1}\bigg(\frac{\partial A^j_j}{\partial q^i}
  - \frac{\partial A^j_i}{\partial q^j} - \frac{\partial A_{ij}}{\partial p_j}
  \bigg)\,\dd q^i\wedge\omega^{n-1}
\nonumber \\
  & & + \frac{1}{n-1}\bigg(\frac{\partial A^{ij}}{\partial q^j}
  + \frac{\partial A^j_j}{\partial p_i} - \frac{\partial A^i_j}{\partial p_j}
  \bigg)\,\dd p_i\wedge\omega^{n-1}
\\
  & = & -\frac{1}{n(n-1)}\,i_{\vect{X}}(\omega^n)
  = \frac{1}{n(n-1)}\,\nu_n(\vect{X}),
\label{nunX}
\end{eqnarray}
where
\begin{equation}
  \vect{X} = \bigg(\frac{\partial A^{ij}}{\partial q^j}
  + \frac{\partial A^j_j}{\partial p_i} - \frac{\partial A^i_j}{\partial p_j}
  \bigg)\,\frac{\partial}{\partial q^i}
  + \bigg(\frac{\partial A_{ij}}{\partial p_j}
  - \frac{\partial A^j_j}{\partial q^i}
  + \frac{\partial A^j_i}{\partial q^j}\bigg)\,
  \frac{\partial}{\partial p_i}
\label{Xexpr} \end{equation} is a volume-preserving vector field
belonging to $\XH^{2n-1}(\mathcal{M},\omega)$. Its relation with
$\phi\alpha$ in eq.~(\ref{phialpha}) is
$$ \phi\alpha = \frac{1}{n(n-1)}\,\vect{X}.
$$

The integral curves of the above volume-preserving vector field
$\vect{X}$ can be obtained by solving the following equations with
the proper initial conditions:
\begin{eqnarray}\nonumber
  \dot{q}^{\,i} & = & \frac{\partial A^{ij}}{\partial q^j}
  + \frac{\partial A^j_j}{\partial p_i}
  - \frac{\partial A^i_j}{\partial p_j}, \\
  \dot{p}_i & = & \frac{\partial A_{ij}}{\partial p_j}
  - \frac{\partial A^j_j}{\partial q^i}
  + \frac{\partial A^j_i}{\partial q^j} .
\label{gHeqs}
\end{eqnarray}
Note that the above equations are nothing but the general
volume-preserving equations on the symplectic manifold
$(\mathcal{M},\omega)$.

\subsection{On The Canonical Hamiltonian Equations,
The Trace of 2-Forms and The Poisson Bracket}

It should be noted that for the case of mechanical system with
\begin{eqnarray}\label{alphaH} \alpha =
\frac{1}{n-1}\,H\,\omega ,\end{eqnarray} where $H$ is a function
on $M$, the general volume-preserving equations (\ref{gHeqs})
become the well known canonical equations and the system becomes a
Hamiltonian system with Hamiltonian function $H$. Therefore, the
equations (\ref{gHeqs}) are the generalization of the canonical
equations in the Hamilton mechanics.

Due to eq.~(\ref{nunX}) and Theorem
\ref{thm:thm6}, the corresponding vector field
$\vect{X}$ belongs to
$\XH^{2n-1}(\mathcal{M},\omega)$ for each 2-form
$\alpha$. On the other hand, given a vector field
  $\vect{X}\in\XH^{2n-1}(\mathcal{M},\omega)$,
there is always an exact $(2n-1)$-form,
$\dd\beta$, say, such that
  $\nu_n(\vect{X}) = \dd\beta$.
Since $\star$ is a linear isomorphism, there
exists a 2-form
  $\alpha = \star\beta$
satisfying eq.~(\ref{nunX}). Therefore, when
$\alpha$ runs over the whole space
$\Omega^2(\mathcal{M})$, the corresponding
$\vect{X}$ runs over the whole space
$\XH^{2n-1}(\mathcal{M},\omega)$. That is, every
  $\vect{X}\in\XH^{2n-1}(\mathcal{M},\omega)$
can be written in the form of eq.~(\ref{Xexpr}).

For the same vector field
$\vect{X}\in\XH^{2n-1}(\mathcal{M},\omega)$, the
2-form $\alpha$ can be chosen at least up to a
closed 2-form. This can be viewed as a symmetry
of eqs.~(\ref{gHeqs}). If we define a function
$\tr\alpha$ as \begin{equation}
  \alpha\wedge\omega^{n-1} = \frac{\tr\alpha}{n} \,\omega^n
\end{equation}
for each 2-form $\alpha$, then, using the formula
\begin{equation}
  \dd p_i\wedge\dd q^j\wedge\omega^{n-1} = \frac{\delta^j_i}{n}\,\omega^n,
\end{equation}
we obtain that
\begin{equation}
  \tr\alpha = A^i_i.
\end{equation}
The above expression is obviously independent of
the choice of the Darboux coordinates. Let
$\vect{X}_{\tr\alpha}$ be the Hamiltonian vector
field corresponding to the function $\tr\alpha$.
Eq.~(\ref{nunX}) indicates that
\begin{equation}
  \vect{X} = \vect{X}_{\tr\alpha} + \vect{X}'
\end{equation}
where $\vect{X}'$ is the extra part on the right
hand side of eq.~(\ref{nunX}). $\vect{X}'$ corresponds to
the traceless part of $\alpha$,
\begin{equation}
  \alpha - \frac{\tr\alpha}{n}\,\omega.
\end{equation}
Hence, for a 2-form $\alpha = \frac{H}{n-1}\,\omega$ with $H$ a
function on $\cal M$, $\tr\alpha = \frac{n}{n-1}\,H$ and the
traceless part of $\alpha$ vanishes.

If $f(q,p)$ is a function on $\mathcal{M}$, then
the derivative
  $\dot{f} = \frac{\dd}{\dd t}f(q(t),p(t))$
satisfies the equation
\begin{equation}
  \dot{f}\,\omega^n = n(n-1)\,\dd\alpha\wedge\dd f\wedge\omega^{n-2}.
\label{dotf}
\end{equation}
In fact, $\dot{f} = (\Lied{X}f)(q,p)$. And,
\begin{displaymath}
  (\Lied{X}f)\,\omega^n = \Lied{X}(f\,\omega^n)
  = \dd (f\,i_{\vect{X}}\omega^n).
\end{displaymath}
Then, according to eq.~(\ref{nunX}),
\begin{eqnarray*}
  (\Lied{X}f)\,\omega^n & = & - n(n-1)\,\dd(f\,\dd\star(\alpha))
  = - n(n-1)\,\dd f\wedge\dd(\alpha\wedge\omega^{n-2})
\\
  & = & - n(n-1)\,\dd f\wedge\dd\alpha\wedge\omega^{n-2}
  = n(n-1)\,\dd\alpha\wedge\dd f\wedge\omega^{n-2}.
\end{eqnarray*}
Thus eq.~(\ref{dotf}) has been proved.

Especially,  as was just mentioned, for the case of $\alpha$
taking value in (\ref{alphaH}), the system (\ref{gHeqs}) turns out
to be the usual Hamiltonian system. Therefore, ${f}$ as an
observable satisfies the canonical equation in terms of the
Poisson bracket:
\begin{displaymath}
  \dot{f} = \{f,H\} := \vect{X}_H f
  = \frac{\partial f}{\partial q^i}\frac{\partial H}{\partial p_i}
  - \frac{\partial f}{\partial p_i}\frac{\partial H}{\partial q^i}.
\end{displaymath}

On the other hand, for such a Hamiltonian system, we can use
eq.~(\ref{dotf}) to obtain
\begin{displaymath}
  \dot{f}\,\omega^n = n\,\dd(H\,\omega)\wedge\dd f\wedge\omega^{n-2}
  = n\,\dd H\wedge\dd f\wedge\omega^{n-1}
  = \tr(\dd H\wedge\dd f)\,\omega^n.
\end{displaymath}
Namely,
\begin{equation}
  \dot{f} = \tr(\dd H\wedge\dd f).
\end{equation}
Thus, we obtain the relation between the Poisson bracket and the
trace of 2-forms:
\begin{equation}
  \{f,H\} = - \tr(\dd f\wedge\dd H).
\end{equation}

\subsection{One Possible Application}

The importance of volume-preserving systems can  be seen from the
following fact:
\begin{thm}
If a system $S$ on $(\mathcal{M},\omega)$ is not volume-preserving, it can be
extended to be a volume-preserving system $S'$ on
$(\mathcal{M}\times\mathbb{R}^2, \omega')$ such that the orbits of $S$ are
precisely the projection of the orbits of $S'$ onto $\mathcal{M}$.
\end{thm}
As a demonstration, let $q^i$ and $p_i$ ($i = 1$, \ldots, $n$) be
the Darboux coordinates on $\mathcal{M}$ and $q^0$, $p_0$ be the
Cartesian coordinates on $\mathbb{R}^2$. Then select $\omega' =
\dd p_\mu\wedge \dd q^\mu = \omega + \dd p_0\wedge\dd q^0$ as the
symplectic structure  on $\mathcal{M}\times\mathbb{R}^2$, where
the  $\mu$ is summed over $0$ to $n$. It should be mentioned that
strictly speaking, $\omega$ in this expression should be
$\pi^*\omega$ in which $\pi: \mathcal{M}\times\mathbb{R}^2
\longrightarrow \mathcal{M}$ is the projection. But, as a
demonstration, we do not try to give a rigorous description.
Suppose that, on $\mathcal{M}$, $\Lied{X}(\omega^n) = D\,
\omega^n$ where
$$\vect{X} = Q^i(q^1,\ldots,q^n,p_1,\ldots,p_n)\,\frac{\partial}{\partial q^i}
 + P_i(q^1,\ldots,q^n,p_1,\ldots,p_n)\,\frac{\partial}{\partial p_i}$$
is the vector field of system $S$, and $D = D(q^i, p_i)$ is a
function. Then a system $S'$ on $\mathcal{M}\times\mathbb{R}^2$ corresponding to
$$\vect{X}' = Q^i\,\frac{\partial}{\partial q^i}
  + P_i\,\frac{\partial}{\partial p_i}
  - D(q^1,\ldots,q^n,p_1,\ldots,p_n)q^0 \,\frac{\partial}{\partial q^0}$$
can be constructed. It can be easily checked that $S'$ is a volume-preserving
system. And the orbits of $S$ are just the projections of the orbits of $S'$.
If all the properties of system $S'$ are known, so are the properties of $S$.

In the previous works \cite{zgw, zgp}, more concrete applications
have been illustrated and the relations have also been studied
between our general form for the volume-preserving systems on
symplectic manifolds and other volume-preserving systems such as
the Nambu mechanics \cite{nambu} and Feng-Shang's
volume-preserving algorithm \cite{FS}. We will not repeat  these
topics here.

\section{Discussions and Conclusions}
\label{sectDC}

In this paper, we have introduced the definition of the
Euler-Lagrange cohomology groups
$\HEL^{2k-1}(\mathcal{M},\omega)$, $1\leqslant k \leqslant n,$ on
symplectic manifolds $(\mathcal{M}^{2n},\omega)$ and studied their
relations with other cohomologies as well as some of their
properties. It is shown that for $k=1, n $,
$\HEL^{1}(\mathcal{M},\omega)$ and
$\HEL^{2n-1}(\mathcal{M},\omega)$ are isomorphic to the de~Rham
cohomology $\HdR^{1}(\mathcal{M})$ and $\HdR^{2n-1}(\mathcal{M})$,
respectively. On the other hand,
$\HEL^{2k-1}(\mathcal{M},\omega)$, $1< k < n,$ is neither
isomorphic to the de~Rham cohomology $\HdR^{2k-1}(\mathcal{M})$
nor to the harmonic cohomology on $(\mathcal{M}^{2n}, \omega)$,
and they are also different from each other in general. To our
knowledge, these Euler-Lagrange cohomology groups on
$(\mathcal{M}^{2n}, \omega)$ have not yet been introduced
systematically before. It is significant to know whether there are
some more important roles played by these cohomology groups to the
symplectic manifolds.

It is also shown that the ordinary canonical equations in Hamilton
mechanics  correspond to 1-forms that  represent trivial element
in   the first Euler-Lagrange cohomology
$\HEL^{1}(\mathcal{M},\omega)$ on the phase space. Analog to this
property, the general volume-preserving equations on phase space
are presented from cohomological point of view with respect to
forms which represent trivial element in the highest
Euler-Lagrange cohomology group
$\HEL^{2n-1}(\mathcal{M},\omega)$. And the ordinary canonical
equations in  Hamilton mechanics become their special cases.  What
about the image parts of other Euler-Lagrange cohomology groups?
Whether they also lead to some dynamical equations? These problems
are still under investigation.

It is well known that there are a series of conservation laws in
classical mechanics closely related to the so-called phase flow of
the canonical equations of the Hamiltonian systems. The
Liouville's theorem \cite{Arnold}, which claims that the volume of
a domain in the phase space is conserved if all points in the
domain move along the phase flow, is just such a representative.
From the Euler-Lagrange cohomological point of view, however,
these conservation laws should be generalized to the
symplectic flow rather than the phase flow. Further, they may be
related directly to the volume-preserving flow. It is well known
that Liouville's theorem plays  very important roles in both the
classical mechanics and statistical physics. It is reasonable to
expect that the generalized versions of Liouville's theorem should
also play some important roles.

We have also introduced the conception of
relative Euler-Lagrange cohomology. It is of
course interesting to see its applications in
Mechanics and Physics.

The first Euler-Lagrange cohomology group  has been introduced in
order to further introduce its time-discrete version in the study
on the discrete mechanics including the symplectic algorithm
\cite{ELcoh2,ELcoh4}. Although the first Euler-Lagrange cohomology
is isomorphic to the first de~Rham cohomology, its time discrete
version is still intriguing  and plays an important role in the
symplectic algorithm. In addition, it has also been introduced in
the field theory and their discrete versions of independent
variables. The latter is closely related to the multi-symplectic
algorithm \cite{ELcoh2,ELcoh4}. It is of course significant to
introduce the higher Euler-Lagrange cohomology groups in these
fields and to explore their applications.

Since the general volume-preserving equations
have been introduced on symplectic manifold from
the cohomological point of view, it is meaningful
to investigate their time-discrete version and
study its relation with the volume-preserving
algorithm.

It is well known that symplectic manifolds are
closely related to the complex manifolds.
Therefore, it is natural to see what role should
be  played by the complex counterparts of the
Euler-Lagrange cohomology groups on the complex
manifolds.

All these topics are under investigation. We will leave some
results on them  for further publications.

\section*{\begin{center}Acknowledgement\end{center}}

We would like to thank Professors
 Z.J.
Shang and  S.K. Wang and Siye Wu for valuable discussions.
Especially, S.K. Wang and Siye Wu have partly joined us and made
contribution to this work.  This work was supported in part by the
National Natural Science Foundation of China (grant Nos. 90103004,
10171096, 19701032, 10071087) and the National Key Project for
Basic Research of China (G1998030601).

\end{document}